\documentstyle[preprint,aps]{revtex}

\begin{document}
\tightenlines
\draft

\title{Enhanced Diffusion and the Continuous Spontaneous 
Localization Model}

\author{L. F. Santos $^1$ and C. O. Escobar $^2$}

\address{$^1$ Departamento de F\'{\i}sica Nuclear \\
Instituto de F\'{\i}sica da Universidade de S\~ao Paulo, 
C.P. 66318, cep 05389-970 \\
S\~ao Paulo, S\~ao Paulo, Brazil\\
lsantos@charme.if.usp.br \\
$^2$ Departamento de  Raios C\'osmicos e Cronologia \\
Instituto de F\'{\i}sica Gleb Wataghin \\
Universidade Estadual de Campinas, C.P. 6165,  cep 13083-970\\
Campinas, S\~ao Paulo, Brazil\\
escobar@ifi.unicamp.br}

\maketitle

\begin{abstract}
We find an analogy between turbulence and 
the dynamics of the continuous spontaneous 
localization model (CSL) of the wave function. 
The use of a standard white noise in the localization 
process gives Richardson's $t^3 $ law for the turbulent diffusion, 
while the introduction of an affine noise in the CSL allows 
us to obtain the intermittency corrections to this law.\\ 
\end{abstract}


\section{Introduction}

In studying the phenomenon of enhanced diffusion, Shlesinger {\it et al} 
\cite{Shlesinger} were led to introduce the concept of a L\'evy walk 
as an extension of the more familiar L\'evy flight \cite{Levy}. 
The basic difference between a L\'evy flight and a L\'evy walk is 
that for the latter, although the walker visits all sites visited 
by the flight, the jumps do not occur instantaneously, but there may be a
time delay before the next jump. Shlesinger {\it et al} 
obtained an integral transport equation involving a 
scaled memory which is nonlocal in space and time. 
Contrary to the infinite mean square displacement 
obtained in a L\'evy flight, the solution of 
such transport equation leads to a finite mean square displacement, 
which has the same time dependence as obtained by Richardson 
($<{\bf R}^{2} , t>\sim t^3 $) in his pioneering studies of turbulence 
\cite{Richardson}. The Mandelbrot intermittency corrections 
\cite{Mandelbrot} are also considered by the authors of reference 
\cite{Shlesinger} and
provide the necessary corrections to Richardson's law, which are 
observed experimentally.

In this paper we show that similar results are obtained in the different 
physical context of the continuous spontaneous localization 
model of quantum mechanics (CSL) \cite{Ghirardi} such 
as  introduced by 
Ghirardi, Pearle and Rimini (GPR) \cite{GPR}.
In this model the wave function is subjected to a 
stochastic process in Hilbert space. In one dimension 
the evolution equation in the Stratonovich form is \cite{tail}

\begin{equation}
d\psi (x,t) =\left\{ [ -iH -\lambda ]dt  +\sqrt{\gamma } \int dq dB(q,t) 
G(x-q) \right\} \psi (x,t),
\end{equation}
where $dB$ is  
a white noise ($<dB(t)> = 0$ and $ <dB(t) dB(0)  >=  
dt$ ) and

\begin{equation}
G(x-q) = \sqrt{\frac{\alpha }{2\pi }} \exp \left[ -\alpha 
\frac{(x-z)^2]}{2}\right] 
\end{equation}
is an indication of the localization of the wave function.
The length parameter $\alpha $  and the frequency parameter 
$\lambda $ are fundamental parameters of the spontaneous reduction 
model developed by
Ghirardi, Rimini and Weber (GRW) \cite{GRW} and 
are related to $\gamma $ according to $\gamma =
\lambda (4\pi /\alpha )^{\frac{1}{2} }$ \cite{Ghirardi,GPR}.
They are chosen in such a way that the 
new evolution equations do not give different results from the usual 
Schr\"odinger unitary evolution for microscopic systems with few degrees 
of freedom, but when a macroscopic system is described 
there is a fast decay of the 
macroscopic linear superpositions which are quickly transformed 
into statistical mixtures \cite{GPR,GRW}.

This analogy between the CSL process and turbulence allows 
us to obtain in section III
the enhanced diffusion, the mean energy input into the turbulent medium,
a Fokker-Planck equation for the probability density in phase space and 
Mandelbrot's intermittency corrections with the introduction of an affine
noise; all in the framework of the beable interpretation of the CSL 
model, which is presented in the next section.

\section{A Beable Interpretation of the CSL Model}

The usual interpretation of quantum mechanics 
deals fundamentally with results 
of measurements and therefore presupposes, besides a 
system, an apparatus 
to perform the measurements. However what the apparatus is and how to 
distinguish it from the system are questions with vague answers. 
In face of 
this problem, Bell \cite{Bell} proposed an interpretation in terms 
of `beables' 
instead of observables. Beables correspond to things that exist 
independently of the observation, therefore they can be 
assigned well defined values. In this 
way we avoid a cut between the microscopic 
(quantum) world and the macroscopic (classical) world.

Vink \cite{Vink} showed that two other well known interpretations of 
quantum mechanics - the causal interpretation associated with Bohm 
\cite{Bohm} and the stochastic interpretation due to Nelson \cite{Nelson} - 
are particular cases of the beable interpretation as developed by Bell. 
Moreover, he proposed that all observables, even those that do not 
commute, can attain beable status simultaneously.

Generalizing Vink's results \cite{Vink}, we have recently 
extended the beable interpretation 
to the GPR model for a free particle. We treated position 
and momentum as beables and showed that in the continuum 
limit they satisfy the following stochastic differential equations 
\cite{nos}

\begin{equation}
dx =\frac{p_{o}}{M} dt 
+ 2\nu  \sqrt{\gamma }  \left[
\int _{0}^{t} \int dq dB(q,t') 
\frac{\partial G(x-q)}{\partial x}\right] dt   + 
(2\nu )^{\frac{1}{2}} dw,
\end{equation}

\begin{equation}
dp = \hbar \sqrt{\frac{\alpha \lambda }{2}} dw.
\end{equation}

In the next section we exploit equations (3) and (4) 
and obtain the main results of this paper.

\section{Turbulence Results}

In equation (3) the first term on the right hand side 
describes a single free particle
deterministic evolution as 
in the de Broglie-Bohm model. The two  
other terms describe the stochastic processes, 
with $dw$ and $dB$ being two independent 
white noises, $<dw(t)> = 0$, $<dw(t) dw(0)> = dt$ 
and $\nu = \hbar /2m$.
The last stochastic term is a 
standard diffusion and the second term, a non standard diffusion 
which exhibits the non-locality of the localization process. This 
second term indicates that the particle position tracks 
the wave function. The position increment induced by this term drives 
the particle to where the wave function is increasing, 
and therefore localizing, according to 
the fluctuating term in equation (1).
Notice that $dw$ and $dB$ are two independent 
white noise, $<dw(t)> = 0$, $<dw(t) dw(0)> = dt$ 
and $\nu = \hbar /2m$. The non standard diffusion term 
is responsible for the $t^3$ behavior for 
the mean square displacement

\begin{equation}
<x^2 (t)> = <x^2 (t)>_{S} + \frac{\alpha \lambda \hbar ^{2} }{6 m^2 } t^3 ,
\end{equation}
where $<x^2 (t)>_{S} $ is the mean square displacement for the free 
Schr\"odinger evolution and the last term corresponds to the 
enhanced diffusion typical of turbulence \footnote{This $t^3$ 
behavior prompted us to investigate a possible analogy with turbulent 
diffusion.}.

With respect to 
the stochastic process for momentum 
we stress the fact that this is a consequence of the 
localization of the wave function, which vanishes when GRW parameters 
($\alpha $, $\lambda $) go to zero. As for the 
fluctuation in momentum, equation (4) gives

\begin{equation}
<p^2 (t)> = <p^2 (t)>_{S} +\frac{\hbar ^2 \alpha \lambda }{2} t.
\end{equation}
where $<p^2 (t)>_{S} $ is the mean square momentum for the Schr\"odinger
evolution.

Working towards a more insightful physical picture of the above
processes, we now consider the mean energy input in a 
turbulent medium and compare with the equivalent quantity in the GRW 
model.

Turbulence theories based on dimensional analysis give 
\cite{Procaccia}

\begin{equation}
<x^2 (t)> \sim  <\epsilon > t^3 ,
\end{equation}
where $<\epsilon >$ is the mean energy input per unit time and 
per unit mass. Comparing (7) with our equation (5), we are led to identify  

\begin{equation}
<\epsilon > =  \frac{\alpha \lambda \hbar ^{2} }{6 m^2 } ,
\end{equation}
which coincides with the term of energy non-conservation 
(per unit time per unit mass) of the GRW collapsing model 
(\cite{GRW}, eqs. 7.1 and 7.2).

A nice feature of our beable interpretation of the CSL model is the 
discontinuous nature of the velocity (eq. 4), which has its own 
analogy in turbulence. This point had already been noticed by 
Richardson \cite{Richardson}, 
although his evolution equation does not take this into account\cite{ref}. 
In our case this comes about naturally as a consequence of giving 
beable status to both position and momentum. From 
the stochastic differential equations (3) and (4), we obtain 
the following Fokker-Planck equation for the probability density
in phase space

\begin{eqnarray}
\frac{\partial P (x,p,t)}{\partial t} &=& 
-\frac{p_{o} }{m} \frac{\partial P(x,p,t) }{\partial x}  \\
                                      &+& \left[ 
\frac{\hbar }{2m} \frac{\partial ^{2} }{\partial x^2 } 
+ \sqrt{\frac{\hbar ^3 \alpha \lambda}{2m}} 
\frac{\partial ^2 }{\partial x \partial p} + 
\frac{\hbar ^2 \alpha \lambda}{4} \frac{\partial ^2 }{\partial p^2 } 
\right] P(x,p,t) \nonumber ,
\end{eqnarray}
which now has two diffusion coefficients for position 
$\hbar /m $ and for momentum 
$\hbar ^2 \alpha \lambda /2$.

We have considered so far only white noise. If we wish to pursue 
the analogies with turbulence even further the corrections for 
intermittency should now  be taken into account. In order to 
do so we consider 
an affine noise \cite{Mandelbrot2} called fractional Brownian noise 

\begin{equation}
< dB (t) dB (0) > = t^{A-1} dt.
\end{equation}
Noise $dB$ when used in equation (3) gives for the anomalous diffusion 
term

\begin{equation}
<x^2 (t)> \sim  t^{A-1+3} .
\end{equation}
This corresponds to one of the intermittency 
corrections obtained by Shlesinger {\it et al} 
provided we identify $A-1$ with
$3 \mu /(4- \mu ) $, where $\mu = E - df $, $E$ being 
the Euclidean dimension and $df$, the fractal dimension.

For the momentum variable the non-white noise gives

\begin{equation}
<p^2 > \sim t^A ,
\end{equation}
which leads to the scaling relation obtained by Shlesinger {\it et al}
\cite{Shlesinger} for the root-mean-square velocity. Notice 
that contrary to reference \cite{Shlesinger}, we obtain 
the intermittency correction (11) without having to use (12).

\section{Conclusion}

We have exploited an analogy between turbulence and the beable 
interpretation of the spontaneous localization model in quantum 
mechanics thus providing an appealing physical 
picture for the localization process. 
The analogy with turbulence led us consider 
a non-white noise for the GPR process, which may be useful in the attempt 
to construct a more realistic model along the lines of GPR 
\cite{tail,pearle}.

Within the beable interpretation of the CSL model, we have found a 
dynamics that has the character of a L\'evy 
dynamics without having to use the L\'evy stable laws. 
This point had already been noticed by Kusnezov {\it et al} 
\cite{Kusnezov}.

\acknowledgments
The authors acknowledge the support of the 
Brazilian Research Council, CNPq.


\begin{references}
 
\bibitem{Shlesinger} M. F. Shlesinger, B. J. West and J. 
Klafter, Phys. Rev. Lett. {\bf 58}, 1100 (1987).
\bibitem{Levy} P. L\'evy, {\it Th\'eorie de 
l'Addition des Variables Al\'eatoires} (Gauthier-Villiers, 1937).
\bibitem{Richardson} L. F. Richardson, Proc. Roy. Soc. London, 
Ser A {\bf 110}, 709 (1926).
\bibitem{Mandelbrot} B. B. Mandelbrot, J. Fluid Mech. {\bf 62}, 
331 (1974).
\bibitem{Ghirardi} G. C. Ghirardi, R. Grassi and F. Benatti, 
Found. Phys. {\bf 25}, 5 (1995).
\bibitem{GPR} G. C. Ghirardi, P. Pearle and  A. Rimini, Phys. Rev. A 
{\bf 42}, 78 (1990).
\bibitem{tail} P. Pearle, {\it Tales and Tails and Stuff and Nonsense}, 
quant-ph/9805050.
\bibitem{GRW} G. C. Ghirardi, A. Rimini and T. Weber, Phys. Rev. D
{\bf 34}, 470 (1986).
\bibitem{Bell} J. S. Bell, Phys. Rep. {\bf 137}, 49 (1986).
\bibitem{Vink} J. C. Vink, Phys. Rev. A {\bf 48}, 1808 (1993).
\bibitem{Bohm} D. Bohm, Phys. Rev. {\bf 85}, 166 (1952), {\bf 85}, 
180 (1952).
\bibitem{Nelson} E. Nelson, Phy. Rev., {\bf 150}, 1079 (1969).
\bibitem{nos} L. F. Santos and C. O. Escobar, {\it A Beable 
Interpretation of the GRW Spontaneous Collapse Model}, quant-ph/9810019.
\bibitem{Procaccia} H. G. E. Hentschel and I. Procaccia, Phys. Rev. A
{\bf 27}, 1266 (1993).
\bibitem{ref} See the clarifying comment made by Shlesinger 
{\it et al} \cite{Shlesinger} on this aspect of Richardson's work.
\bibitem{Mandelbrot2} B. B. Mandelbrot {\it Fractals Form, Chance, and 
Dimension} (W. H. Freeman and Company, 1977).
\bibitem{pearle} P. Pearle, Phys. Rev. A {\bf 48}, 913 (1993).
\bibitem{Kusnezov} D. Kusnezov, A. Bulgac and  G. D. Dang, Phys. Lett. 
{\bf A234}, 103 (1997).

\end{references}
\end{document}